\documentclass[journal,10pt]{IEEEtranTCOM}

\normalsize

\usepackage[T1]{fontenc}
\usepackage{amsfonts}
\usepackage{amsmath}
\usepackage{amsthm}
\usepackage{amssymb}
\usepackage{caption}
\usepackage[labelformat=simple]{subcaption}
\usepackage{cite,graphicx,algorithm}
\usepackage{algorithmic}
\usepackage{color}
\usepackage{bm}

\ifCLASSINFOpdf
\else
\fi

\theoremstyle{plain}

\begin{document}
%
\title{Linear Precoding Design for OTFS Systems in Time/Frequency Selective Fading Channels}
%
%
%

\author{Yao~Ge,~\IEEEmembership{Member,~IEEE,}
        Lingsheng~Meng,~\IEEEmembership{Student Member,~IEEE,}
        David~Gonz\'{a}lez~G.,~\IEEEmembership{Senior Member,~IEEE,}
        Miaowen~Wen,~\IEEEmembership{Senior Member,~IEEE,}
        Yong~Liang~Guan,~\IEEEmembership{Senior Member,~IEEE,}
        and~Pingzhi~Fan,~\IEEEmembership{Fellow,~IEEE}
\thanks{This study is supported under the RIE2020 Industry Alignment Fund—Industry Collaboration Projects (IAF-ICP) Funding Initiative, as well as cash and in-kind contribution from the industry partner(s).}
\thanks{Yao Ge, Lingsheng Meng, and Yong Liang Guan are with the Continental-NTU Corporate Lab, Nanyang Technological University, Singapore 639798 (e-mail: yao.ge@ntu.edu.sg; meng0071@e.ntu.edu.sg; eylguan@ntu.edu.sg).}
\thanks{David~Gonz\'{a}lez~G. is with the Wireless Communications Technologies Group, Continental AG, 65936 Frankfurt am Main, Germany (e-mail: david.gonzalez.g@ieee.org).}
\thanks{Miaowen Wen is with the School of Electronic and Information Engineering, South China University of Technology, Guangzhou 510641, China (e-mail: eemwwen@scut.edu.cn).}
\thanks{Pingzhi Fan is with the Key Laboratory of Information Coding and Transmission of Sichuan Province, CSNMT International Cooperation Research Centre (MoST), Southwest Jiaotong University, Chengdu 610031, China (e-mail: p.fan@ieee.org).}
}

%
%

\markboth{}%
{}
%



\maketitle

\begin{abstract}
Even orthogonal time frequency space (OTFS) has been shown as a promising modulation scheme for high mobility doubly-selective fading channels, its attainability of full diversity order in either time or frequency selective fading channels has not been clarified. By performing pairwise error probability (PEP) analysis, we observe that the original OTFS system can not always guarantee full exploitation of the embedded diversity in either time or frequency selective fading channels. To address this issue and further improve system performance, this work proposes linear precoding solutions based on algebraic number theory for OTFS systems over time and frequency selective fading channels, respectively. The proposed linear precoded OTFS systems can guarantee the maximal diversity and potential coding gains in time/frequency selective fading channels without any transmission rate loss and do not require the channel state information (CSI) at the transmitter. Simulation results are finally provided to illustrate the superiority of our proposed precoded OTFS over both the original unprecoded and the existing phase rotation OTFS systems in time/frequency selective fading channels.
\end{abstract}

\begin{IEEEkeywords}
Diversity gain, time/frequency selective fading channels, linear precoding, OTFS, PEP analysis.
\end{IEEEkeywords}

%
\IEEEpeerreviewmaketitle

\section{Introduction}
%
%
%
%

High data rates and multipath propagation give rise to frequency-selectivity of wireless channels, while carrier frequency offsets (CFOs) and Doppler caused by mobility between the transmitter and receiver introduce time-selectivity in wireless channels \cite{tse2005fundamentals}. Orthogonal frequency division multiplexing (OFDM) is particularly attractive in practice because it can transform a frequency-selective fading channel into parallel flat-fading sub-channels with the use of a sufficiently long cyclic prefix (CP) \cite{7469313}. However, the performance of uncoded OFDM degrades significantly as it can not exploit the multipath diversity, and guarantee the orthogonality among subcarriers in Doppler time-selective fading channels.

A linear constellation precoded OFDM system is proposed in \cite{1194447} for multicarrier transmissions over multipath frequency-selective fading channels without an essential decrease in transmission rate. Meanwhile, a space-time-Doppler coded system is developed in \cite{1433146} that guarantees the maximum possible space-Doppler diversity, along with the largest coding gains in time-selective fading channels. In \cite{1207384}, a block precoded transmission is proposed for single-carrier communications to guarantee the maximum diversity gain in doubly-selective fading channels. However, to achieve the full diversity and avoid inter-block interference, a CP/zero padding (ZP) guard interval is inserted per block at the transmitter and discarded at the receiver \cite{1194447,1433146} and the spreading technique is applied in \cite{1207384}, leading to lower spectral efficiency caused by the more significant CPs/ZPs or lower spreading gain. 

Recently, orthogonal time frequency space (OTFS) modulation \cite{hadani2017orthogonal} has been proposed as a promising and alternative PHY-layer modulation scheme to traditional OFDM for high mobility communications. Only one CP is required for the whole OTFS frame, leading to high spectral efficiency compared to traditional OFDM systems.
The diversity performance analysis of uncoded and coded OTFS systems have been respectively analyzed and evaluated in \cite{surabhi2019diversity,raviteja2019effective} and \cite{li2021performance} over high-mobility doubly-selective fading channels. However, attainability of the OTFS full diversity order in time/frequency selective fading channels was not clarified nor has been proved theoretically in the literature. By performing pairwise error probability (PEP) analysis, we observe that the original OTFS system cannot always guarantee full exploitation of the embedded diversity in time/frequency selective fading channels. Therefore, it is of interest to develop efficient methods for OTFS systems that can guarantee both performance and high spectral efficiency in time/frequency selective fading channels.

In this work, we propose linear precoding schemes for OTFS systems based on algebraic number theory, which effectively realizes the maximal diversity and potential coding gains in time/frequency selective fading channels. It turns out that the proposed linear precoding matrix can be verified to guarantee the maximum diversity order irrespective of the system dimension, and without any transmission rate loss. The performance merits of our precoding design are confirmed by corroborating simulations and compared with original unprecoded and the existing phase rotation OTFS systems.

\section{System Model}\label{II_model}

\begin{figure}
  \centering
  \includegraphics[width=3.5in]{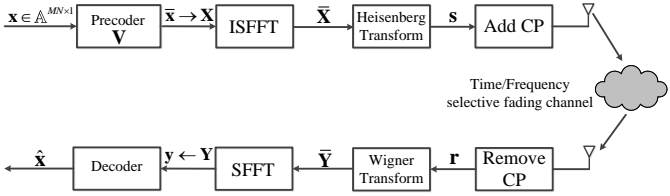}
  \caption{Block diagram of the proposed precoded OTFS system for time/frequency selective fading channels.}
  \label{fig1_diagram}
  \vspace{-1.5em}
\end{figure}
As shown in Fig. \ref{fig1_diagram}, we consider a precoded OTFS transmission over time/frequency selective fading channels.
 
\subsection{Transmitter}
Without loss of generality, the information streams ${\bf{x}} \in {\mathbb{A}^{MN \times 1}}$ are drawn from a finite modulation alphabet $\mathbb{A}$ (e.g., phase shift keying (PSK) and quadrature amplitude modulation (QAM) symbols), where $M$ and $N$ represent the numbers of resource grids along the OTFS delay and Doppler dimensions, respectively. After linear precoding, we can obtain the transmitted OTFS symbols ${\bf{\bar x}} \in {\mathbb{C}^{MN \times 1}}$ as
\begin{align}
{\bf{\bar x}} = {\bf{Vx}},
\end{align}
where ${\bf{V}} \in {\mathbb{C}^{MN \times MN}}$ is the precoding matrix and will be designed later on to guarantee the maximum diversity gain. 

We then arrange the information symbols ${\bf{\bar x}} \in {\mathbb{C}^{MN \times 1}}$ into the two-dimensional delay-Doppler plane ${\bf{X}} \in {\mathbb{C}^{M \times N}}$, i.e., ${\bf{X}} = \text{invec}({\bf{\bar x}})$. By first applying the inverse symplectic finite Fourier transform (ISFFT) on ${\bf{X}}$ followed by Heisenberg transform with a rectangular transmit pulse, the resulted output can be given by
\begin{align}
{\bf{S}} = {\bf{F}}_M^H\left( {{{\bf{F}}_M}{\bf{XF}}_N^H} \right) = {\bf{XF}}_N^H,
\end{align}
where ${{\bf{F}}_M} \in {\mathbb{C}^{M \times M}}$ and ${{\bf{F}}_N} \in {\mathbb{C}^{N \times N}}$ are the normalized $M$-point and $N$-point fast Fourier transform (FFT) matrices, respectively. The transmitted time domain signal ${\bf{s}} \in {\mathbb{C}^{MN \times 1}}$ is then generated by column-wise vectorization of ${\bf{S}}$.

To overcome the inter-frame interference, we add a CP in front of the generated time domain signal with a length no shorter than the maximal channel delay spread. The resulted time domain signal is finally sent to the receiver through the channel.

\subsection{Channel Model}

\textbf{Multipath frequency-selective fading channel:} High data rates and multipath propagation give rise to frequency-selectivity of wireless channels. Here, we characterize the multipath frequency-selective fading channel as a finite-impulse response ${\bf{h}} = {\left[ {h[0],h[1], \cdots ,h[L - 1]} \right]^T} \in {\mathbb{C}^{L \times 1}}$, where ${h[p]}$ is the complex gain for the $p$-th channel tap with $p\in\{0,1,\ldots, L-1\}$, and $L$ is the maximum number of channel taps. 

\textbf{High-mobility time-selective fading channel:} Carrier frequency offsets and Doppler caused by the mobility between the transmitter and receiver lead to time-selectivity of wireless channels. Basis expansion model (BEM) has been widely adopted to parameterize the time
varying channel as a weighted combination of basis functions \cite{1433146}. The baseband channel impulse response can be characterized as
\begin{align}
h[c] = \sum\limits_{q = 0}^Q {{c_q}{e^{j{\omega _q}c}}},
\end{align}
where $Q = 2\left\lceil {N{{\bar f}_{\max }}} \right\rceil$ is the order of BEM basis functions, ${{\bar f}_{\max }} = {{{f_{\max }}} \mathord{\left/
 {\vphantom {{{f_{\max }}} {\Delta f}}} \right.
 \kern-\nulldelimiterspace} {\Delta f}}$ with ${{f_{\max }}}$ being the maximum Doppler frequency and $\Delta f$ being the subcarrier interval. ${\omega _q} = \frac{{2\pi }}{{MN}}\left( {q - \left\lceil {\frac{Q}{2}} \right\rceil } \right)$ denotes the $q$-th BEM modeling frequency and ${{c_q}}$ is the $q$-th BEM channel coefficient. $\left\lceil  \cdot  \right\rceil$ represents the round up operation. Here, the channel $h[c]$ changes along with time index $c$ and the Doppler spread is controlled by the maximum Doppler frequency ${{f_{\max }}}$, i.e., the Doppler spread may consist of multiple Doppler shifts which are no larger than the maximum Doppler frequency. 
Note that the BEM facilitates our development and analysis of the diversity for OTFS systems over time-selective fading channels.

\subsection{Receiver}
At OTFS receiver, we can obtain the received signal ${\bf{r}} \in {\mathbb{C}^{MN \times 1}}$ after removing CP as
\begin{align}
r[c] \!=\! \sum\limits_{p = 0}^{L - 1} {h[p]s\left[ {{{[c \!-\! p]}_{MN}}} \right]} \! +\! n[c],c \!= \!0,1, \cdots ,MN - 1
\end{align}
for frequency-selective fading channels and 
\begin{align}
r[c] = h[c]s[c]  + n[c],\;c = 0,1, \cdots ,MN - 1
\end{align}
for time-selective fading channels.
${\bf{n}} \in {\mathbb{C}^{MN \times 1}} \sim \mathcal{CN}\left( {{\bf{0}},{N_0}{\bf{I}}} \right)$ is the received noise and the notation ${[ \cdot ]_m}$ denotes mod-$m$ operation.

The received time domain signal ${\bf{r}}\in {\mathbb{C}^{MN \times 1}}$ is then devectorized into a matrix ${\bf{R}} \in {\mathbb{C}^{M \times N}}$, followed by Winger transform with a rectangular receive pulse as well as the symplectic finite Fourier transform (SFFT), yielding the recovered delay-Doppler domain signal
\begin{align}
{\bf{Y}} = {\bf{F}}_M^H\left( {{{\bf{F}}_M}{\bf{R}}} \right){{\bf{F}}_N} = {\bf{R}}{{\bf{F}}_N}.
\end{align}

\vspace{-0.3cm}
\section{Performance Analysis and Precoding Design}

In this section, we derive the performance criteria for the precoded OTFS systems, and also determine the maximum achievable diversity gain and analyze the corresponding coding gain for both time and frequency selective fading channels. 

\subsection{Frequency Selective Fading Channel Scenario}

For frequency-selective fading channels, the end-to-end input-output relationship of OTFS transmission in delay-Doppler domain can be vectorized column-wise into \cite{ge2021otfs}
\begin{subequations}\label{IO_Multipath}
\begin{align}
{\bf{y}} &\!=\! \left(\! {{{\bf{F}}_N} \!\otimes \!{{\bf{I}}_M}}\!\right)\!{\bf{F}}_{MN}^H\text{diag}\!\left\{ {{{\bf{F}}_{MN \times L}}{\bf{h}}} \right\}\!{{\bf{F}}_{MN}}\!\left(\! {{\bf{F}}_N^H \!\otimes\! {{\bf{I}}_M}} \!\right)\!{\bf{\bar x}}\\
& = {\bf{H}}{\bf{Vx}}
 = {\bf{\Phi }}\left( {\bf{x}} \right){\bf{h}},
\end{align}
\end{subequations}
where ${\bf{H}} \!=\! \left( {{{\bf{F}}_N} \!\otimes \!{{\bf{I}}_M}} \right)\!{\bf{F}}_{MN}^H\text{diag}\!\left\{ {{{\bf{F}}_{MN \!\times \!L}}{\bf{h}}} \right\}\!{{\bf{F}}_{MN}}\left( {{\bf{F}}_N^H \!\otimes \!{{\bf{I}}_M}} \right)$ and ${\bf{\Phi }}\!\left( \!{\bf{x}}\! \right)\! =\! \left(\! {{{\bf{F}}_N} \!\otimes\! {{\bf{I}}_M}}\! \right)\!{\bf{F}}_{MN}^H\text{diag}\!\left\{\! {{{\bf{F}}_{MN}}\!\left(\! {{\bf{F}}_N^H \!\otimes\! {{\bf{I}}_M}}\! \right)\!{\bf{Vx}}}\! \right\}\!{{\bf{F}}_{MN \!\times\! L}}$. Note that we omit the noise term in (\ref{IO_Multipath}) for notational brevity.

Assuming perfect channel state information (CSI) is available at the receiver, the conditional
PEP, i.e., the probability of transmitting ${\bf{x}}$ but erroneously
deciding on ${{\bf{\hat x}}}$, is given by
\begin{align}\label{condit_PEP}
\Pr \left\{ {\left. {{\bf{x}} \to {\bf{\hat x}}} \right|{\bf{h}}} \right\} = Q\left( {\sqrt {\frac{\rho }{2}{{\left\| {\left( {{\bf{\Phi }}({\bf{x}}) - {\bf{\Phi }}({\bf{\hat x}})} \right){\bf{h}}} \right\|}^2}} } \right),
\end{align}
where $Q\left( x \right)$ is the tail distribution function of the standard Gaussian distribution and $\rho  = \frac{1}{{{N_0}}}$ denotes the signal-to-noise ratio (SNR).

Note that ${\bf{C}} = {\left( {{\bf{\Phi }}({\bf{x}}) - {\bf{\Phi }}({\bf{\hat x}})} \right)^H}\left( {{\bf{\Phi }}({\bf{x}}) - {\bf{\Phi }}({\bf{\hat x}})} \right)$ is a Hermitian matrix, its rank and the non-zero eigenvalues are defined as $R$ and ${\lambda _i},i = 1,2, \cdots ,R$, respectively. Hence, we can obtain 
\begin{align}
{\left\| {\left(\! {{\bf{\Phi }}({\bf{x}})\! -\! {\bf{\Phi }}({\bf{\hat x}})} \!\right)\!{\bf{h}}} \right\|^2} \!\!=\! {{\bf{h}}^H}\!{\bf{U\Sigma }}{{\bf{U}}^H}\!{\bf{h}}
\!=\! {{{\bf{\bar h}}}^H}\!{\bf{\Sigma \bar h}}
\!=\!\! \sum\limits_{i = 1}^R \!{{\lambda _i}{{\left| {{{\bar h}_i}} \right|}^2}}\label{Hermi_deco},
\end{align}
where ${\bf{U}}\in {\mathbb{C}^{L \times L}}$ is a unitary matrix, ${\bf{\bar h}} = {{\bf{U}}^H}{\bf{h}}$ and ${\bf{\Sigma }} = \text{diag}\left\{ {{\lambda _1},{\lambda _2}, \cdots ,{\lambda _L}} \right\}$.

Substituting (\ref{Hermi_deco}) in (\ref{condit_PEP}), the conditional PEP is rewritten as
\begin{align}
\Pr \!\left\{ {\left. \!{{\bf{x}} \!\to\! {\bf{\hat x}}} \right|\!{\bf{h}}} \!\right\} \!=\! Q\!\left(\!\! {\sqrt {\frac{\rho }{2}\!\sum\limits_{i = 1}^R \!{{\lambda _i}{{\left| {{{\bar h}_i}} \right|}^2}} } } \right)
 \!\le\! \frac{1}{2}\!\prod\limits_{i = 1}^R \!{\exp \left( \!{ - \frac{\rho }{4}{\lambda _i}{{\left| {{{\bar h}_i}} \right|}^2}} \right)}.\label{condi_PEP_appro} 
\end{align}

Since ${{\bf{\bar h}}}$ is obtained by multiplying a unitary matrix with ${\bf{h}}$, it has the same distribution as that of ${\bf{h}}$. 
The elements in ${{\bf{\bar h}}}$ are assumed to be independent and identically distributed complex Gaussian random variables. 
Considering ${\bf{\bar h}} \sim \mathcal{CN}\left( {{\bf{0}},\frac{1}{L}{\bf{I}}} \right)$, the final PEP is calculated by averaging (\ref{condi_PEP_appro}) over the channel statistics and given by
\begin{align}
\Pr \left\{ {{\bf{x}} \!\to \!{\bf{\hat x}}} \right\} \!=\! \mathbb{E}\!\left[ {Q\left( {\sqrt {\frac{\rho }{2}\sum\limits_{i = 1}^R {{\lambda _i}{{\left| {{{\bar h}_i}} \right|}^2}} } } \right)} \right]
 \!\le \!\frac{1}{2}\prod\limits_{i = 1}^R {\frac{1}{{1 \!+\! \frac{\rho }{4}\frac{{{\lambda _i}}}{L}}}}, \label{PEP_high}
\end{align}
where $\mathbb{E}\left[  \cdot  \right]$ represents the expectation operation. At high SNRs (i.e., $\rho  \to \infty $), (\ref{PEP_high}) can be further simplified as
\begin{align}
\Pr \!\left\{ {{\bf{x}} \!\to\! {\bf{\hat x}}} \right\} \!\le\! \frac{1}{2}{\left(\! {\prod\limits_{i = 1}^R {\frac{{{\lambda _i}}}{{4L}}} } \!\right)^{ - 1}}{\rho ^{ - R}}
 \!=\! \frac{1}{2}{\left(\! {\frac{{{{\left( {\prod\limits_{i = 1}^R {{\lambda _i}} } \right)}^{\frac{1}{R}}}}}{{4L}}} \!\right)^{ - R}}\!{\rho ^{ - R}}.\nonumber
\end{align}

From the above analysis, we conclude that the system diversity order is determined by $R$, which could be as high as the number of resolvable paths $L$ of the channel. The term ${{{\left( {\prod\limits_{i = 1}^R {{\lambda _i}} } \right)}^{\frac{1}{R}}}}$ stands for the pairwise coding gain to control how this PEP shifts relative to the benchmark error-rate curve of ${\left( {{\rho  \mathord{\left/
 {\vphantom {\rho  {4L}}} \right.
 \kern-\nulldelimiterspace} {4L}}} \right)^{ - R}}$. Accounting for all possible pairwise errors, we define herein the diversity and coding gains, respectively, as 
\begin{align}\label{diver_coding_define}
{G_d} = \mathop {\min }\limits_{{\bf{x}} \ne {\bf{\hat x}}} R, \quad\quad {G_c} = \mathop {\min }\limits_{{\bf{x}} \ne {\bf{\hat x}}} {\left( {\prod\limits_{i = 1}^R {{\lambda _i}} } \right)^{\frac{1}{R}}}.
\end{align}
Because the system performance depends on both ${G_d}$ and ${G_c}$, it is important to maximize both ${G_d}$ and ${G_c}$. By checking the dimensionality of ${\bf{C}}$, it is clear that the maximum diversity gain ${G_{d,\max }} = L$ is achieved if and only if the matrix ${\bf{C}}$ has full rank (i.e., $\det \left( {\bf{C}} \right) \ne 0$) $\forall {\bf{x}} \ne {\bf{\hat x}}$. When the maximum diversity gain ${G_{d,\max }} = L$ is achieved, the coding gain becomes 
\begin{align}\label{coding}
{G_c} = \mathop {\min }\limits_{{\bf{x}} \ne {\bf{\hat x}}} \det {\left( {{{\bf{R}}_h}} \right)^{\frac{1}{L}}}\det {\left( {\bf{C}} \right)^{\frac{1}{L}}},
\end{align}
where ${{\bf{R}}_h} = \mathbb{E}\left[ {{\bf{h}}{{\bf{h}}^H}} \right]$. Equation (\ref{coding}) implies that ${G_c}$ is a function of the determinant
\begin{align}
&\det \left( {\bf{C}} \right) = \det \left( {{{\left( {{\bf{\Phi }}({\bf{x}}) - {\bf{\Phi }}({\bf{\hat x}})} \right)}^H}\left( {{\bf{\Phi }}({\bf{x}}) - {\bf{\Phi }}({\bf{\hat x}})} \right)} \right)\nonumber\\
&\!=\!\det \!\left( \!{{{\left( {\text{diag}\!\left\{ \!{{\bf{\Theta }}\!\left( {{\bf{x}}\! -\! {\bf{\hat x}}} \right)}\! \right\}{{\bf{F}}_{MN \!\times \!L}}} \right)}^H}\!\!\left( {\text{diag}\!\left\{ \!{{\bf{\Theta }}\!\left( {{\bf{x}}\! -\! {\bf{\hat x}}} \right)} \!\right\}{{\bf{F}}_{MN\! \times\! L}}} \right)}\! \right)\nonumber\\
&\!=\! \prod\limits_{j = 1}^L \!{{\lambda _j}\!\left( {\text{diag}\!\left\{\! {{\bf{\Theta }}\!\left( {{\bf{x}} \!- \!{\bf{\hat x}}} \right)}\! \right\}{{\bf{F}}_{MN \times L}}{\bf{F}}_{MN \times L}^H \text{diag}{{\left\{ \!{{\bf{\Theta }}\!\left( {{\bf{x}} \!-\! {\bf{\hat x}}} \right)} \!\right\}}^H}} \right)} \nonumber\\
& = \prod\limits_{j = 1}^L {{\beta _j}{\lambda _j}\left( {{{\bf{F}}_{MN \times L}}{\bf{F}}_{MN \times L}^H} \right)} \nonumber\\
&= \prod\limits_{j = 1}^L {{\beta _j}}  \times \det \left( {{\bf{F}}_{MN \times L}^H{{\bf{F}}_{MN \times L}}} \right),
\end{align}
where ${\bf{\Theta }} = {{\bf{F}}_{MN}}\left( {{\bf{F}}_N^H \otimes {{\bf{I}}_M}} \right){\bf{V}}$ and ${{\bm{\theta }}_i^T}$ is the $i$-th row of ${\bf{\Theta }}$. ${\lambda _i}\left( {\bf{A}} \right)$ is the $i$-th non-zero eigenvalue of matrix ${\bf{A}}$ and $0 < \mathop {\min }\limits_{i \in 1,2, \cdots ,MN} {\left| {{\bm{\theta }}_i^T\left( {{\bf{x}} - {\bf{\hat x}}} \right)} \right|^2} \le {\beta _j} \le \mathop {\max }\limits_{i \in 1,2, \cdots ,MN} {\left| {{\bm{\theta }}_i^T\left( {{\bf{x}} - {\bf{\hat x}}} \right)} \right|^2}$. The last equality follows from the Ostrowski’s theorem \cite{1512125}. As ${{{\bf{F}}_{MN \times L}}}$ is the first $L$ principal columns of $MN$-point FFT matrix, $\det \left( {{\bf{F}}_{MN \times L}^H{{\bf{F}}_{MN \times L}}} \right) = {\left( {MN} \right)^L}$.  

Certainly, the diversity gain ${G_d}$ and the coding gain ${G_c}$ are both depend on the choice of ${\bf{V}}$. Without a proper precoding matrix ${\bf{V}}$, one can not achieve the potential diversity and coding gains, leading to a significant performance loss. At high SNR, it is reasonable to maximize the diversity gain first, because it determines the slope of the log-log bit-error rate (BER)-SNR curve. Note that ${{{\bf{F}}_{MN \times L}}}$ is full rank. We can guarantee that the matrix ${\bf{C}}$ has full rank if $\text{diag}\left\{ {{\bf{\Theta }}\left( {{\bf{x}} - {\bf{\hat x}}} \right)} \right\}$ is also full rank $\forall {\bf{x}} \ne {\bf{\hat x}}$. Interestingly, a class of important Vandermonde/unitary matrix ${\bf{\Theta }}$ is proposed in \cite{1512125,1194447} and constructed by using the algebraic number theory for MIMO and OFDM systems. Here, we set ${\bf{\Theta }}$ as a Vandermonde matrix 
\begin{align}\label{theat_begin}
\setlength{\arraycolsep}{1.2pt}
{\bf{\Theta }} = \frac{1}{\xi }\left[ {\begin{array}{*{20}{c}}
1&{{\alpha _1}}& \cdots &{\alpha _1^{MN - 1}}\\
1&{{\alpha _2}}& \cdots &{\alpha _2^{MN - 1}}\\
 \vdots & \vdots & \ddots & \vdots \\
1&{{\alpha _{MN}}}& \cdots &{\alpha _{MN}^{MN - 1}}
\end{array}} \right],
\end{align}
where $\xi $ is a normalization factor chosen to guarantee the power constraint $\text{Tr}\left( {{\bf{V}}{{\bf{V}}^H}} \right) = MN$, and the selection of parameters 
$\left\{ {{\alpha _k}} \right\}_{k = 1}^{MN}$ depends on $MN$, for example:

If $MN = {2^d}$ ($d \ge 1$), the ${\alpha _k}$ is determined as 
\begin{align}
{\alpha _k} = {e^{j\frac{{4k - 3}}{{2MN}}\pi }},k = 1,2, \cdots ,MN.
\end{align}

If $MN = 3 \times {2^d}$ ($d \ge 0$), the ${\alpha _k}$ is specified as
\begin{align}
{\alpha _k} = {e^{j\frac{{6k - 1}}{{3MN}}\pi }},k = 1,2, \cdots ,MN.
\end{align}

If $MN = {2^d} \times {3^t}$ ($d \ge 1,t \ge 1$), the ${\alpha _k}$ is given by
\begin{align}\label{theat_end}
{\alpha _k} = {e^{j\frac{{6k - 5}}{{3MN}}\pi }},k = 1,2, \cdots ,MN.
\end{align}

For more details and other cases of $MN$, one can refer to \cite{1512125,1512140}. After obtaining ${\bf{\Theta }}$, the precoding matrix\footnote{Note that only FFT process is involved, making the proposed precoder relatively easy to
implement in practice.} is given by
\begin{align}
{\bf{V}} = \left( {{{\bf{F}}_N} \otimes {{\bf{I}}_M}} \right){\bf{F}}_{MN}^H{\bf{\Theta }}
\end{align}
to achieve the maximum diversity gain of OTFS systems in frequency-selective fading channels. The corresponding coding gain is characterized as 
\begin{align}
{G_c} = \mathop {\min }\limits_{{\bf{x}} \ne {\bf{\hat x}}} MN \times {\left( {\det \left( {{{\bf{R}}_h}} \right)\prod\limits_{j = 1}^L {{\beta _j}} } \right)^{\frac{1}{L}}},
\end{align}
where $0 \!<\! \mathop {\min }\limits_{i \in 1,2, \cdots ,MN} {\left| {{\bm{\theta }}_i^T\left( {{\bf{x}} - {\bf{\hat x}}} \right)} \right|^2} \!\le\! {\beta _j} \!\le\! \mathop {\max }\limits_{i \in 1,2, \cdots ,MN} {\left| {{\bm{\theta }}_i^T\left( {{\bf{x}} - {\bf{\hat x}}} \right)} \right|^2}$.
\vspace{-1em}

\subsection{Time Selective Fading Channel Scenario}

For time-selective fading channels, the end-to-end input-output relationship of OTFS transmission in delay-Doppler domain is vectorized column-wise given by \cite{ge2021otfs}
\begin{subequations}\label{IO_Doppler}
\begin{align}
{\bf{y}} &\!=\! \sum\limits_{q = 0}^Q \!{\left(\! {{{\bf{F}}_N} \!\otimes\! {{\bf{I}}_M}} \!\right)\!{{\bf{D}}_q}{\bf{F}}_{MN}^H\text{diag}\!\left\{ {{{\bf{F}}_{MN \!\times\! 1}}{c_q}} \right\}\!{{\bf{F}}_{MN}}\!\left( \!{{\bf{F}}_N^H \!\otimes\! {{\bf{I}}_M}}\! \right)\!{\bf{\bar x}}}\\
& = \sum\limits_{q = 0}^Q {{{\bf{\Phi }}_q}({\bf{x}}){c_q}}
= {\bf{\Phi }}\left( {\bf{x}} \right){\bf{h}},
\end{align}
\end{subequations}
where ${{\bf{D}}_q} = \text{diag}\left\{ {1,{e^{j{\omega _q}}},{e^{j2{\omega _q}}}, \cdots ,{e^{j{\omega _q}(MN - 1)}}} \right\}$, ${{\bf{\Phi }}_q}({\bf{x}}) = \left( {{{\bf{F}}_N} \otimes {{\bf{I}}_M}} \right){{\bf{D}}_q}\left( {{\bf{F}}_N^H \otimes {{\bf{I}}_M}} \right){\bf{Vx}}$ and ${\bf{h}} \in {\mathbb{C}^{(Q + 1) \times 1}} = {\left[ {{c_0},{c_1}, \cdots ,{c_Q}} \right]^T}$. We also express ${\bf{\Phi }}\left( {\bf{x}} \right) \in {\mathbb{C}^{MN \times (Q + 1)}}$ as
\begin{align}
{\bf{\Phi }}\left( {\bf{x}} \right) &= \left[ {{{\bf{\Phi }}_0}({\bf{x}}),{{\bf{\Phi }}_1}({\bf{x}}), \cdots ,{{\bf{\Phi }}_Q}({\bf{x}})} \right]\nonumber\\
&= \left( {{{\bf{F}}_N} \otimes {{\bf{I}}_M}} \right)\text{diag}\left\{ {\left( {{\bf{F}}_N^H \otimes {{\bf{I}}_M}} \right){\bf{Vx}}} \right\}{\bf{B}},
\end{align}
where ${\bf{B}} = \left[ {{{\bf{b}}_0},{{\bf{b}}_1}, \cdots ,{{\bf{b}}_Q}} \right]$ with ${{\bf{b}}_q} = {\left[ {1,{e^{j{\omega _q}}},{e^{j2{\omega _q}}}, \cdots ,{e^{j{\omega _q}(MN - 1)}}} \right]^T}$. Similarly, we omit the noise term in (\ref{IO_Doppler}) for notational brevity.

Considering a unitary matrix ${\bf{U}}\in {\mathbb{C}^{(Q+1) \times (Q+1)}}$ and defining ${\bf{\bar h}}= {{\bf{U}}^H}{\bf{h}} \sim \mathcal{CN}\left( {{\bf{0}},\frac{1}{Q+1}{\bf{I}}} \right)$, the final PEP is calculated similar to (\ref{condit_PEP})-(\ref{PEP_high}), and given by
\begin{align}
\Pr \left\{ {{\bf{x}} \to {\bf{\hat x}}} \right\} \le \frac{1}{2}\prod\limits_{i = 1}^R {\frac{1}{{1 + \frac{\rho }{4}\frac{{{\lambda _i}}}{Q+1}}}}. \label{PEP_high_D}
\end{align}
At high SNRs (i.e., $\rho  \to \infty $), (\ref{PEP_high_D}) can be further simplified as
\begin{align}
\Pr \!\left\{ \!{{\bf{x}} \!\to\! {\bf{\hat x}}} \!\right\} \!\le\! \frac{1}{2}\!{\left(\! {\prod\limits_{i = 1}^R {\frac{{{\lambda _i}}}{{4(Q\!+\!1)}}} } \!\!\right)^{ - 1}}\!{\rho ^{ - R}}
\! =\! \frac{1}{2}\!{\left(\!\! {\frac{{{{\left(\! {\prod\limits_{i = 1}^R {{\lambda _i}} } \!\right)}^{\frac{1}{R}}}}}{{4(Q\!+\!1)}}} \!\!\right)^{ - R}}\!\!{\rho ^{ - R}}.\nonumber
\end{align}

From the above analysis, we conclude that the system diversity order is determined by $R$, which could be as high as the number of bases $Q+1$ in the BEM. Accounting for all possible pairwise errors, the diversity and coding gains are defined similar to (\ref{diver_coding_define}).  
By checking the dimensionality of ${\bf{C}}$, it is clear that the maximum diversity gain ${G_{d,\max }} = Q+1$ is achieved if and only if the matrix ${\bf{C}}$ has full rank (i.e., $\det \left( {\bf{C}} \right) \ne 0$) $\forall {\bf{x}} \ne {\bf{\hat x}}$. When the maximum diversity gain ${G_{d,\max }} = Q+1$ is achieved, the coding gain becomes 
\begin{align}\label{coding_D}
{G_c} = \mathop {\min }\limits_{{\bf{x}} \ne {\bf{\hat x}}} \det {\left( {{{\bf{R}}_h}} \right)^{\frac{1}{Q+1}}}\det {\left( {\bf{C}} \right)^{\frac{1}{Q+1}}},
\end{align}
where ${{\bf{R}}_h} = \mathbb{E}\left[ {{\bf{h}}{{\bf{h}}^H}} \right]$. Equation (\ref{coding_D}) implies that ${G_c}$ is a function of the determinant
\begin{align}
&\det \left( {\bf{C}} \right) = \det \left( {{{\left( {{\bf{\Phi }}({\bf{x}}) - {\bf{\Phi }}({\bf{\hat x}})} \right)}^H}\left( {{\bf{\Phi }}({\bf{x}}) - {\bf{\Phi }}({\bf{\hat x}})} \right)} \right)\nonumber\\
&\!=\!\det \left( {{{\left( {\text{diag}\left\{ {{\bf{\Theta }}\left( {{\bf{x}} - {\bf{\hat x}}} \right)} \right\}{\bf{B}}} \right)}^H}\left( {\text{diag}\left\{ {{\bf{\Theta }}\left( {{\bf{x}} - {\bf{\hat x}}} \right)} \right\}{\bf{B}}} \right)} \right)\nonumber\\
&\!=\! \prod\limits_{j = 1}^{Q + 1} {{\lambda _j}\left( {\text{diag}\left\{ {{\bf{\Theta }}\left( {{\bf{x}} - {\bf{\hat x}}} \right)} \right\}{\bf{B}}{{\bf{B}}^H}\text{diag}{{\left\{ {{\bf{\Theta }}\left( {{\bf{x}} - {\bf{\hat x}}} \right)} \right\}}^H}} \right)} \nonumber\\
&= \prod\limits_{j = 1}^{Q + 1} {{\beta _j}}  \times \det \left( {{{\bf{B}}^H}{\bf{B}}} \right),
\end{align}
where ${\bf{\Theta }} = \left( {{\bf{F}}_N^H \otimes {{\bf{I}}_M}} \right){\bf{V}}$ and ${{\bm{\theta }}_i^T}$ is the $i$-th row of ${\bf{\Theta }}$. $0 < \mathop {\min }\limits_{i \in 1,2, \cdots ,MN} {\left| {{\bm{\theta }}_i^T\left( {{\bf{x}} - {\bf{\hat x}}} \right)} \right|^2} \le {\beta _j} \le \mathop {\max }\limits_{i \in 1,2, \cdots ,MN} {\left| {{\bm{\theta }}_i^T\left( {{\bf{x}} - {\bf{\hat x}}} \right)} \right|^2}$.
Note that ${\bf{B}} \in {\mathbb{C}^{MN \times (Q + 1)}}$ is full rank. We can guarantee that the matrix ${\bf{C}}$ has full rank if $\text{diag}\left\{ {{\bf{\Theta }}\left( {{\bf{x}} - {\bf{\hat x}}} \right)} \right\}$ is also full rank for $\forall {\bf{x}} \ne {\bf{\hat x}}$. The choice of ${\bf{\Theta }}$ is similar to (\ref{theat_begin})-(\ref{theat_end}). After obtaining ${\bf{\Theta }}$, the precoding matrix is given by
\begin{align}
{\bf{V}} = \left( {{{\bf{F}}_N} \otimes {{\bf{I}}_M}} \right){\bf{\Theta }}
\end{align}
to achieve the maximum diversity gain of OTFS systems in time-selective fading channels. The corresponding coding gain is characterized as 
\begin{align}
{G_c} = \mathop {\min }\limits_{{\bf{x}} \ne {\bf{\hat x}}} {\left( {\prod\limits_{j = 1}^{Q + 1} {{\beta _j}} \det \left( {{{\bf{B}}^H}{\bf{B}}} \right)\det \left( {{{\bf{R}}_h}} \right)} \right)^{\frac{1}{{Q + 1}}}},
\end{align}
where $0 \!<\! \mathop {\min }\limits_{i \in 1,2, \cdots ,MN} {\left| {{\bm{\theta }}_i^T\left( {{\bf{x}} - {\bf{\hat x}}} \right)} \right|^2} \!\le\! {\beta _j} \!\le\! \mathop {\max }\limits_{i \in 1,2, \cdots ,MN} {\left| {{\bm{\theta }}_i^T\left( {{\bf{x}} - {\bf{\hat x}}} \right)} \right|^2}$.
\vspace{-1em}

\section{Simulation Results}\label{V_simulation}
\begin{figure}
  \centering
  \includegraphics[width=2.6in]{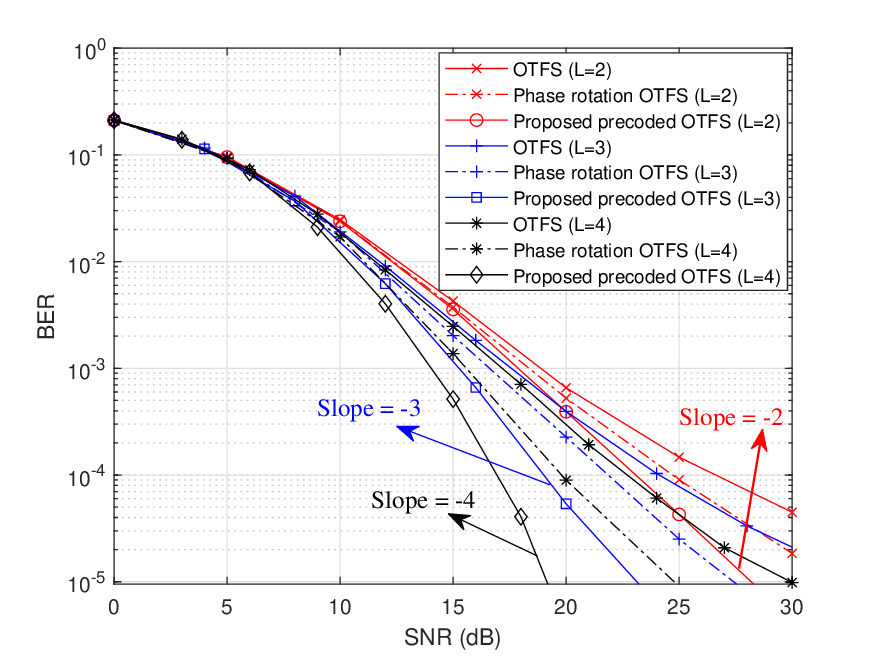}
  \caption{BER performance comparison with different number of resolvable paths under ML detector.}\label{ML_detector_compare}
  \vspace{-1.5em}
\end{figure}
In this section, we test the performance of our proposed precoded OTFS systems for time and frequency selective fading channels, respectively. We consider the carrier frequency is centered at $4$ GHz and subcarrier spacing $\Delta f=15$ kHz. 
We assume that the perfect channel knowledge is known at the receiver and apply QPSK modulation.

\begin{figure*}
\begin{minipage}[t]{0.33\textwidth}
\centering
\includegraphics[width=1\linewidth]{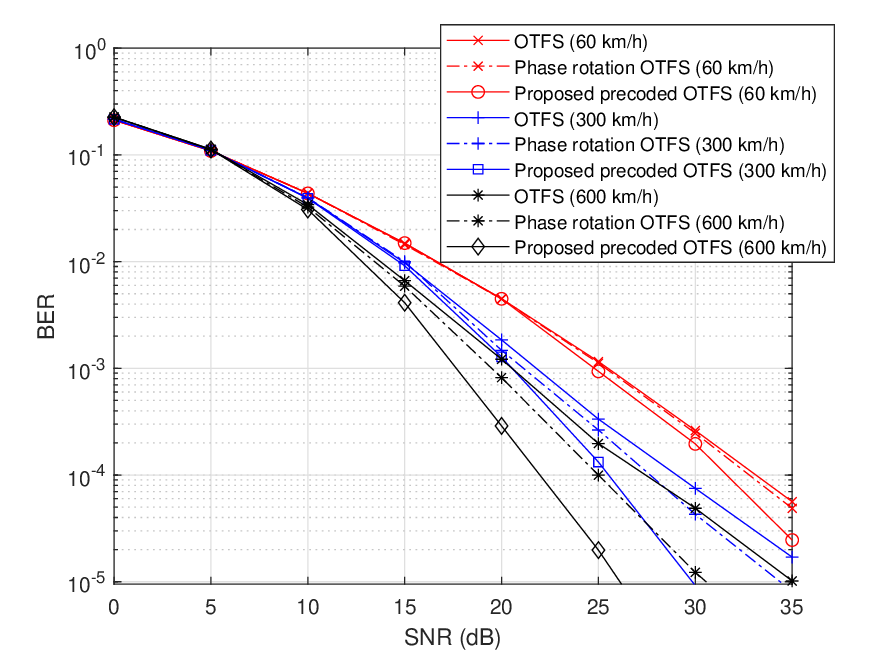}
\caption{BER performance comparison with different user velocities under ML detector.}\label{ML_detector_compare_D}
\end{minipage}
\begin{minipage}[t]{0.33\textwidth}
\centering
\includegraphics[width=1\linewidth]{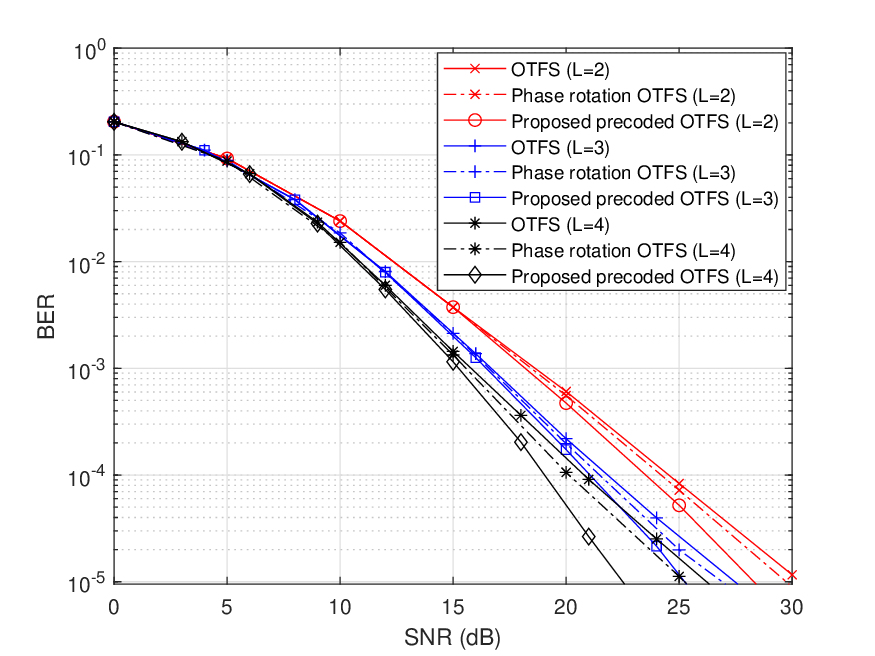}
\caption{BER performance comparison with different number of resolvable paths under Memory AMP.}\label{MAMP_detector_compare}
\end{minipage}
\begin{minipage}[t]{0.33\textwidth}
\centering
\includegraphics[width=1\linewidth]{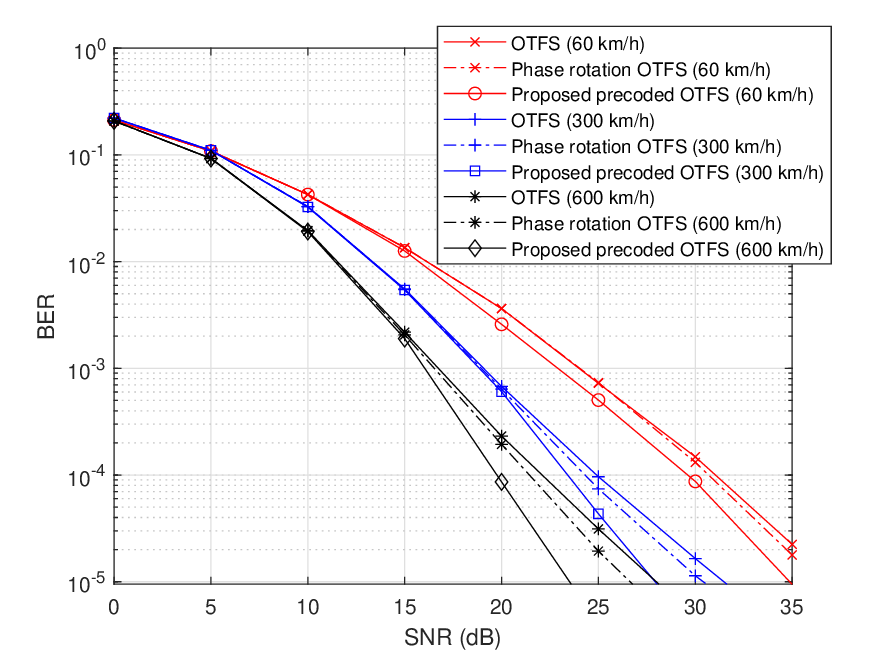}
\caption{BER performance comparison with different user velocities under Memory AMP.}\label{MAMP_detector_compare_D}
\end{minipage}
\vspace{-1.5em}
\end{figure*}
We first examine the effectiveness of the proposed precoding results for OTFS systems with maximum likelihood (ML) detector. Fig. \ref{ML_detector_compare} shows the BER performance for different number of resolvable paths (i.e., frequency-selective fading channel) with a delay-Doppler plane $M=4$ and $N=2$. It is obvious that the BER performance improves as the number of resolvable paths $L$ increases for both the precoded, unprecoded and existing phase rotation \cite{surabhi2019diversity} OTFS systems. This is due to the fact that a high diversity gain can be obtained for better performance with a large value of $L$. We also notice that our proposed precoded OTFS system outperforms the traditional unprecoded and phase rotation ones, and can achieve the maximal diversity gain in the multipath frequency selective fading channels.

Fig. \ref{ML_detector_compare_D} further illustrates the BER performance for different user velocities (i.e., time-selective fading channel) with a delay-Doppler plane $M=2$ and $N=4$. It is obvious that the BER performance improves as the user velocity increases for both the precoded, unprecoded and existing phase rotation \cite{surabhi2019diversity} OTFS systems. This is due to the fact that a high Doppler diversity gain can be obtained for better performance with a large value of user velocity. We also notice that our proposed precoded OTFS system outperforms the traditional unprecoded and phase rotation ones, and can achieve the potential maximal diversity gain to improve the system performance.

As the complexity of ML detector grows exponentially with the system dimension, it cannot be directly applied to practical large dimensional systems due to intolerable computational burden. We now test the BER performance in large dimension systems, where the practical low complexity advanced Memory approximate message passing (AMP) detector \cite{10283752} is applied to further verify the advantage of our proposed precoding results for OTFS systems compared to the traditional unprecoded and phase rotation ones. From the results in Fig. \ref{MAMP_detector_compare} ($M=128$ and $N=16$), we can observe that the BER performance of both precoded, unprecoded and existing phase rotation OTFS systems improve as $L$ increases since the potential higher diversity can be exploited from a larger number of independent resolvable paths. Our proposed precoded OTFS system still outperforms the traditional unprecoded and phase rotation ones by using the practical low complexity detectors.  

Similarly, Fig. \ref{MAMP_detector_compare_D} presents the BER performance for different
user velocities with a delay-Doppler plane $M=128$ and $N=16$ by using the practical low complexity Memory AMP detector. From the results in Fig. \ref{MAMP_detector_compare_D}, we can observe that the BER performance of both precoded, unprecoded and existing phase rotation OTFS systems improve as user velocity increases. Our proposed precoded OTFS system still outperforms the traditional unprecoded and phase rotation ones in such time selective fading channels. 


\section{Conclusion}\label{VI_conclusion}
In this work, we proposed linear precoding schemes for OTFS system based on algebraic number theory. The PEP analysis verified that our proposed precoded OTFS system can achieve the maximal diversity and potential coding gains for wireless transmissions over time/frequency selective fading channels. The proposed precoding design for OTFS does not require the CSI at the transmitter and can be used for an arbitrary system dimension without any transmission rate loss. Our results demonstrated that the proposed precoding design for OTFS system exhibits sufficient statistic diversity of time/frequency selective fading channels, and outperforms the original unprecoded and existing phase rotation OTFS systems for both optimal ML detector and low-complexity advanced Memory AMP detector.


%





\ifCLASSOPTIONcaptionsoff
  \newpage
\fi



%
%




\vspace{-1em}
\bibliographystyle{IEEEtran}
\footnotesize
\bibliography{ref_OTFS_CoMP}

\begin{thebibliography}{10}
\providecommand{\url}[1]{#1}
\csname url@samestyle\endcsname
\providecommand{\newblock}{\relax}
\providecommand{\bibinfo}[2]{#2}
\providecommand{\BIBentrySTDinterwordspacing}{\spaceskip=0pt\relax}
\providecommand{\BIBentryALTinterwordstretchfactor}{4}
\providecommand{\BIBentryALTinterwordspacing}{\spaceskip=\fontdimen2\font plus
\BIBentryALTinterwordstretchfactor\fontdimen3\font minus
  \fontdimen4\font\relax}
\providecommand{\BIBforeignlanguage}[2]{{%
\expandafter\ifx\csname l@#1\endcsname\relax
\typeout{** WARNING: IEEEtran.bst: No hyphenation pattern has been}%
\typeout{** loaded for the language `#1'. Using the pattern for}%
\typeout{** the default language instead.}%
\else
\language=\csname l@#1\endcsname
\fi
#2}}
\providecommand{\BIBdecl}{\relax}
\BIBdecl

\bibitem{tse2005fundamentals}
D.~Tse and P.~Viswanath, \emph{Fundamentals of {W}ireless
  {C}ommunication}.\hskip 1em plus 0.5em minus 0.4em\relax Cambridge university
  press, 2005.

\bibitem{7469313}
B.~Farhang-Boroujeny and H.~Moradi, ``{OFDM} inspired waveforms for {5G},''
  \emph{IEEE Commun. Surv. Tuts.}, vol.~18, no.~4, pp. 2474--2492,
  Fourthquarter 2016.

\bibitem{1194447}
Z.~Liu, Y.~Xin, and G.~Giannakis, ``Linear constellation precoding for {OFDM}
  with maximum multipath diversity and coding gains,'' \emph{IEEE Trans.
  Commun.}, vol.~51, no.~3, pp. 416--427, Mar. 2003.

\bibitem{1433146}
X.~Ma, G.~Leus, and G.~Giannakis, ``Space-time-{D}oppler block coding for
  correlated time-selective fading channels,'' \emph{IEEE Trans. Signal
  Process.}, vol.~53, no.~6, pp. 2167--2181, Jun. 2005.

\bibitem{1207384}
X.~Ma and G.~Giannakis, ``Maximum-diversity transmissions over doubly selective
  wireless channels,'' \emph{IEEE Trans. Inf. Theory}, vol.~49, no.~7, pp.
  1832--1840, Jul. 2003.

\bibitem{hadani2017orthogonal}
R.~Hadani \emph{et~al.}, ``Orthogonal time frequency space modulation,'' in
  \emph{Proc. IEEE Wireless Commun. Netw. Conf. (WCNC)}, San Francisco, CA,
  USA, Mar. 2017, pp. 1--6.

\bibitem{surabhi2019diversity}
G.~Surabhi, R.~M. Augustine, and A.~Chockalingam, ``On the diversity of uncoded
  {OTFS} modulation in doubly-dispersive channels,'' \emph{IEEE Trans. Wireless
  Commun.}, vol.~18, no.~6, pp. 3049--3063, Jun. 2019.

\bibitem{raviteja2019effective}
P.~Raviteja, Y.~Hong, E.~Viterbo, and E.~Biglieri, ``Effective diversity of
  {OTFS} modulation,'' \emph{IEEE Wireless Commun. Lett.}, vol.~9, no.~2, pp.
  249--253, Feb. 2020.

\bibitem{li2021performance}
S.~Li, J.~Yuan, W.~Yuan, Z.~Wei, B.~Bai, and D.~W.~K. Ng, ``Performance
  analysis of coded {OTFS} systems over high-mobility channels,'' \emph{IEEE
  Trans. Wireless Commun.}, vol.~20, no.~9, pp. 6033--6048, Sep. 2021.

\bibitem{ge2021otfs}
Y.~Ge, Q.~Deng, P.~Ching, and Z.~Ding, ``{OTFS} signaling for uplink {NOMA} of
  heterogeneous mobility users,'' \emph{IEEE Trans. Commun.}, vol.~69, no.~5,
  pp. 3147--3161, May 2021.

\bibitem{1512125}
L.~Shao and S.~Roy, ``Rate-one space-frequency block codes with maximum
  diversity for {MIMO}-{OFDM},'' \emph{IEEE Trans. Wireless Commun.}, vol.~4,
  no.~4, pp. 1674--1687, Jul. 2005.

\bibitem{1512140}
W.~Su, Z.~Safar, and K.~R. Liu, ``Towards maximum achievable diversity in
  space, time, and frequency: performance analysis and code design,''
  \emph{IEEE Trans. Wireless Commun.}, vol.~4, no.~4, pp. 1847--1857, Jul.
  2005.

\bibitem{10283752}
Y.~Ge \emph{et~al.}, ``Low-complexity memory {AMP} detector for high-mobility
  {MIMO}-{OTFS} {SCMA} systems,'' in \emph{IEEE International Conference on
  Communications Workshops (ICC Workshops)}, May 2023, pp. 807--812.

\end{thebibliography}

%




\end{document}